%% file: molec.tex
\documentclass[10pt,twoside,draft,twocolumn]{article}
\usepackage{ulem}
\usepackage[dvips]{graphics}
\usepackage{epsf}
\usepackage{color}
\begin{document}
\title{Thermodynamics of Information Retrieval}
%\numberofauthors{3}
%\author{K. Koroutchev}
%\affiliation{
%  Escuela Polit\'ecnica Superior,
%  Universidad Aut\'onoma de Madrid, \\
%  28049 Canto Blanco, Madrid, Spain
%\email{Corresponding author, tel: +34-91-497-3210, fax: +34-91-497-22-35,\\
%k.koroutchev@uam.es}
%}
%\author{E. Korutcheva}\altaffiliation[Also at ]{G.Nadjakov Inst. Solid State
% Physics, %\\
% Bulgarian Academy of Sciences, Sofia, Bulgaria
%}
%\affiliation{Depto. F\'{\i}sica Fundamental, UNED\\
% c/Senda del Rey No 9, 28080 Madrid, Spain
%}
\date{}
\author{Kostadin Koroutchev\thanks{K.K. is with the Department of Computer Sicence at Autonomous University of Madrid, 28049 Madrid, Spain (E-mail: k.koroutchev@uam.es).}, Jian Shen\thanks{J.S. is with the Department of Computer Sicence at Autonomous University of Madrid, 28049 Madrid, Spain (E-mail: jian.shen@estudiante.uam.es).}, Elka Koroutcheva\thanks{E.K. is with the department of Fundamental Physics at National Distance Education University, 28080 Madrid, Spain (E-mail: elka@fisfun.uned.es). She is also with the  G. Nadjakov Institute for Solid State Physics, Bulgarian Academy of Sciences, 1784 Soﬁa, Bulgaria.} and Manuel Cebri\'an\thanks{M.C. is with the department of Computer Science at Brown University, RI 02912, USA (E-mail: mcebrian@cs.brown.edu).}}

%\author{
%% e-mail address with \email.
% \alignauthor Kostadin Koroutchev and Jian Shen\\
%        \affaddr{Autonomous University of Madrid}\\
%        \affaddr{28049 Madrid, Spain}\\
%        \email{k.koroutchev@uam.es}\\
%        \email{jian.shen@estudiante.uam.es}
% \alignauthor Elka Korutcheva\titlenote{Elka Koroutcheva is also at
%  G. Nadjakov Inst. Solid State Physics, Bulgarian Academy of Sciences, 1784 Soﬁa, Bulgaria.}\\
%        \affaddr{Dept. of Fundamental Physics}\\
%        \affaddr{UNED}\\
%         \affaddr{28080 Madrid, Spain}\\
%        \email{elka@fisfun.uned.es} 
% \alignauthor Manuel Cebri\'an\\
%        \affaddr{Dept. of Computer Science}\\
%        \affaddr{Brown University}\\
%        \affaddr{RI 02912, USA}\\
%        \email{mcebrian@cs.brown.edu}
% }
%\pacs{64.10.+h, 64.60.De, 65.40.-b, 89.75.-k}

\maketitle
\input{00_abstract.tex}

% \category{E.4}{Data}{Coding and Information Theory}[Formal models of
% communication]
% \category{G.3}{Mathematics of Computing}{Probability and Statistics}
% \category{H.3.3}{Information Systems}{Information Storage and Retrieval}[Information Search and Retrieval]

% \terms{Languages, Theory}

% \keywords{Keywords, Natural Language, Grammar, Entropy, Energy,
% Statistical Mechanics}

%\renewcommand{\baselinestretch}{2}
\input{01_intro.tex}
\input{02_method.tex}
\input{03_single_word.tex}
\input{04_free_energy.tex}
\input{05_numerical.tex}

\input{06_application.tex}
\input{07_discussion.tex}
\input{08_relatedwork.tex}
\input{09_conclusion.tex}
\section*{Acknowledgments:}
The authors thank the International Center for Theoretical
Physics, Trieste, Italy, where this investigation has been
completed. Especially we would like to thank the Statistical
Physics group at ICTP.
% for stimulating discussions.
The work is
financially supported by Grants TIN 2004-07676-G01-01 (K.K.) and
DGI.M.CyT FIS2005-1729 (E.K.) from the Spanish Ministry of Science
and Education.

\end{document}

%% file: 00_abstract.tex
\begin{abstract}

In this work, we suggest a parameterized statistical model (the gamma
distribution) for the frequency of word occurrences in long strings of
english text and use this model to build a corresponding thermodynamic
picture by constructing the partition function. We then use our
partition
 function to compute
thermodynamic quantities such as the free energy and the specific heat.
In this approach, the parameters of the word frequency model vary from
word to word so that each word has a different corresponding
thermodynamics and we suggest that differences in the specific
heat reflect differences in how the words are used in language,
differentiating
 keywords
 from common and function words.  Finally, we apply our thermodynamic
picture to the problem of retrieval of texts based on keywords and
suggest some advantages over traditional information retrieval methods.

%In this article we present preliminary ideas of text search by means of a model of human written text based on
%statistical mechanics consideration. The text is conditioned to the
%language in which it is written, which is represented by a large
%text corpus. As a first approximation we use the words as elementary
%units composing the text which interact with the language.
%By using general principles from the statistical physics, we derive
%empirically the potential energy for the different parts of the text
%and we calculate the thermodynamic parameters of the system.
%It worth mention that the adopted approach allows us to derive a closed analytical form for the %solution.
%The results have been checked on a set of short texts (scientific
%articles and novels) and it has been shown that the ``specific
%heat'' parameter effectively separates the function words from the
%specific terms (keywords) used in the text. Finally we give an
%example of a possible application of the method in information
%retrieval and a brief discussion
%of some future directions.
\end{abstract}

%% file: 01_intro.tex
 \section{Introduction}

 Let us imagine that we are looking for some article in the
 Web. Probably the first thing we will do is to go to a search engine and type some keywords.
 If we type a query like  ``I am looking for an article about statistical mechanics of images'', although it is exactly what we want, we will probably
 get nothing related to the subject or we will get only a content partially related to it.  In order to have some meaningful results, one needs to
 refine the query to something like ``image'', ``statistical mechanics'' ignoring in this way the structure of the language and using some statistical
 estimations of the parts of the query that stick well with its meaning.

 Current web search engines are a product of some 15 years evolution. This evolution has shown that if we are  looking for the meaning of a text,
 we must look for specific, statistically salient keywords that are supposed  to be present in it, largely ignoring the syntactic and the semantics
 structure of the language.

Probably, the best way to do the analysis of a text,  written is
some language \cite{footnote1}, would be to have some exact
descriptions of the language, for example, a weighted  context-free
grammar \cite{grammar}. Having in mind the Zipf's law \cite{zipf} of
the frequency distribution of the words, even if reasonable grammar
exists,  in a single text of arbitrary length we will have some
40\% halomorphemes \cite{footnote2}.  As a consequence, the length
of the grammar will be of the order of  the length of the text   for
any text we choose.

 Therefore, it is convenient to consider the language as a set of all the texts spoken/written  in that language.
 Using statistical arguments, we do not need all texts, but only a significantly  large random set of texts in order to
 treat the problem.

In this article we propose a statistical physics model of the text that treats the text as a large random data set.
The text is regarded to be conditioned on the language in which the  text is written and can be restricted on the area
to which it belongs, as for example ``nonlinear physics''  or ``novels of 17$^{\rm th}$ century''.

 The model we investigate consists of a text $T$ and a vocabulary  $V$, written in some language.
 The vocabulary is formed using as a basis some huge collection of texts,  written in that language.

 The relationship between the vocabulary and the text is asymmetric.
 If we regard an article of nonlinear science, it is highly probable to find  words
 like ``chaotic dynamics'' or ``Hamiltonian'', but highly improbable to  find words
 like ``horses'' and ``knights''. Regarding, for example ``Don Quixote'', it is just the opposite.
 So a text that treats some subject is highly restricted  by this subject and the later conditions the vocabulary used.
 The language as a whole has no such restriction. Therefore, the relative excess  (or higher frequency)
 of a word in the vocabulary is a normal situation.

 On the contrary, the relative excess of a word in the text has a specific meaning,  because if the
 word  is with much higher occurrence in the text than in the common language, that can  be  interpreted
 as an indication that this text treats exactly a subject expressed by  this  word, e.g. that the word is a
 {\it specific term} or {\it keyword} in the text.  This is the first class of words in the text that we
 will consider in this article.

 On the other hand, the text will always contain words that are common in the  language, which have more or
 less the same frequency in any text and in the  vocabulary. A large fraction of the words of that type will
 be formed by the so called  {\it function words}.  These words by themselves carry no meaning but are essential
 for expressing the  language structure.  A typical example of a function word in English is the word ``the''.
 The problem with this category is that it is not very easy to define it in a way  that can be implemented by a
 computer program. A similar and strictly defined category is the class of closed class words that  by definition
 are the words, which do not change their form in any text.

 Finally the third class of words that will follow more or less the same  frequency distribution in the text
 and in the vocabulary are the  {\it common words}.  They serve to transmit the meaning of the text, but are
 common for every text that must explain some concept, like for example the word ``explain''  in this sentence.
 In this class significant deviations between different texts and different  authors can be expected.

 In the literature, the statistical treatment of the text is mainly regarded in relation with the
 information retrieval (IR) theory, where this consideration  results very fruitful \cite{baeza,nlp_ir}.

 Another statistical consideration is centered on the Zipf law
\cite{zipf, baayen} and looks for the
 relative distribution of different words (types) in a collection of texts. The Zipf law can be derived
 from the requirement of maximal information exchange \cite{zipf2}.  This approach mainly focuses on the
 tail of the distribution, that is an example of large number of rare events (LNRE).

 In this article we fix the length of the text to some reasonable value (10000 words) and consider
 it in relation to some dictionary. Having fixed number of words in consideration, we do note have to regard the LNRE
 type of distribution.

The main contributions of this article are:

\begin{itemize}
\item The gamma distribution is a better model of word occurrences
  then other models considered in the literature.

\item  The specific heats of different words reflect important differences
 in how words are used in language.

\item The thermodynamic picture offers advantages when searching for
relevant texts based on a set of keywords.
 
\end{itemize}

 The paper is organized in the following way: In the Section 2, we define the model and the approximations used.
 In Section 3 we derive an expression of the frequency of a given word
 in a fixed length text and the
 potential energy corresponding to this probability distribution in the thermodynamic limit.
 In Section 4 we derive an analytical expression for the free energy of a text and the corresponding
 thermodynamics quantities.  Using the results from Section 4, in Section 5 we calculate numerically
 these thermodynamic quantities for a set of arbitrary selected texts and we find that the specific terms (keywords)
 and the rest of the text have different thermodynamic
 behavior. Section 7  presents our discussion
 and comments about the future directions of the work. Section 8
 briefly summarizes the research related the present work, and section 9 presents the
 conclusions of the article.

%% file: 02_method.tex
 \section{The Model}
 In our approach we use the following metaphor to explain the model. We consider the vocabulary as a solid-state basement,
 composed by ``molecules'', which form the parts of the text.
 The text itself is considered as a liquid solution of ``molecules'', derived in the same manner as the vocabulary.
 The text and the vocabulary ``react'' and there exists some energy gain when the reaction takes place, so some
 ``molecules'' are settled down on the solid base.

 As a first approximation, the molecules can be assumed to react only if they represent one and the same word in the
 text and the vocabulary.  A typical text has insignificant length compared to the vocabulary and practically the words
 of the text will ``deposit'', except the orthographic errors, the words defined in the text and probably the foreign
 proper names. To have a consideration of the text almost independent of its length, we can impose the requirement to
 have equal total number of ``molecules'' in the solid and the liquid phase. This can be achieved by replicating the
 text the times necessary to achieve one and the same length of the text and the vocabulary.

 Our model thus consists of a vocabulary of length $L_v$, a text of length $L_t$, and the ``molecules'' (words)
 of the text $w$ that match to the ``molecules'' of the vocabulary. The corresponding number of occurrences
 of these ``molecules''are $n_t(w)$ and $n_v(w)$ for the text and for the vocabulary, respectively.  In order to fulfill the requirement of
 equal length between the text and the vocabulary, we can introduce some standard text length $L_0$ and normalize the
 number of occurrence of $w$ according to this
 length: \[ N_t(w)=L_0\frac{n_t(w)}{L_t},\ \ \ N_v(w)=L_0\frac{n_v(w)}{L_v}. \] For convenience we choose $L_0=L_t$ in
 the numerical experiments. We denote the number of deposited molecules, normalized to length $L_0$ by $m(w)$.
 This parameter will be used below as an order parameter for   the system.

 The problem of regarding the text as a thermodynamic system consists of defining the ``molecules'' $w$ and
 the energy of the interaction  $E(w)=E(m(w),N_t(w),N_v(w),L_0)$ between the language and the text. In this article we will
 regard as ``molecules'' the usual English words, consisting of continuous strings of letters, separated by non-letter
 symbols in written texts.  In the rest of the article we will not distinguish between ``molecules'' and words.
 As a first approximation we assume that the words are independent, e.g. that there is no interaction between different
 words. Due to this assumed independence, the extensive thermodynamics quantities, as for example the free energy,
 will be the sum of the corresponding quantities over the words.  Therefore, we can build a theory, based on a single
 word and extrapolate it on the text.

 Further, we consider that the language (the solid compound) imposes some potential energy field with strength dependent
 on the $N_v$, $L_v$ but not on the text, e.g. not on $N_t$ (when it is not required we will omit the $w$ argument).
 We also assume that the system is in thermal equilibrium.

 According to this consideration, the probability $P(m)$ of the state with $m$ deposited molecules is \cite{Beck}:
\begin{equation}
P(m) \propto G(m) \exp(- \beta E(m, N_t, N_v, L_0)),\label{pw}
 \end{equation}
where $E(m, N_t,N_v, L_0)$ is the energy of settling $m$ molecules,
$G(m, N_t)$ is the number of degenerations of these states and
$\beta$ is the inverse temperature $\beta\equiv1/T$. The number of
degenerations is just the number of ways we can select $m$
``molecules'' out of a set of $N_t$ molecules, e.g. $G(m, N_t) =
{{N_t}\choose {m}}$. Note that this number is strictly zero if
$m>N_t$, that reflects the fact that we have only $N_t$ molecules.

 Regarding that system, one can impose the requirement that its properties scale with the length of
 texts e.g. if we scale simultaneously the size of the vocabulary and the text by $s$, the thermodynamics potential
 will scale in the following way: \[ E(s m, s N_t, s N_v, s L_0)= s E(m, N_t, N_v, L_0) \]
 and \[ \log(G(s m, s N_t))= s \log(G(m, N_t)). \] This requirements must to be fulfilled only in the asymptotic
 limit, e.g. when $s\rightarrow\infty$, which permits to use the saddle point approximation in Eq.~\ref{pw},
 considering as important only the limit

 \[ \lim_{s\rightarrow\infty} [E(s m, s N_t, s L_t, s N_v, s L_v)\beta+\log(G(s m))]/s. \]

%% file: 03_single_word.tex
\section{Frequency of a single word}

 Let us consider the frequency of occurrence of a single word $w$ in a text with length $L$,
 regarding just the case where the word occurs $x\gg 1$ times. The question is what is the probability
 distribution of a given word in this segment of text.  The answer can be given only by empirical argument
 investigating a large repository of texts.

 The usual hypothesis is that the distribution is binomial or mixture of Binomials that corresponds to some urn
 process \cite{baayen}. More sophisticated models suppose that the distribution is a mixture of Binomial
 (when the word is not used as a keyword) and a Flat distribution (when the word is used as a keyword)
 \cite{mixbinom,labbe1,labbe2}. Some process is assumed to be responsible of this distribution, where the
 probability of having the word in a text increases if the word is already in the text. This leads to a mixture of
 Poisson processes.

\begin{figure}[t]
\begin{center}
\begin{minipage}{7.5cm}
\epsfxsize 7.5cm %\epsfysize=5cm
\epsfbox{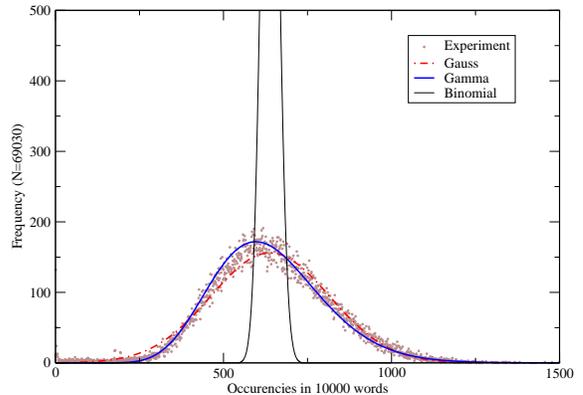}
\end{minipage}
\caption{\small (Color online) Frequency distribution of the word
``the'' in 10000 consecutive words of the corpus. The dots
represent the empirical data; the red line the best Poisson/Normal
distribution fit and the blue line -- the best Gamma fit. The
black line is the binomial distribution that corresponds to the
empirical parameters. }\label{the_distr_3.eps}
\end{center}
 \end{figure}

However, we have found that the distribution is far from Binomial.
As an illustration, in Fig. \ref{the_distr_3.eps} we give the frequency
distribution of the word ``the'' in the Gutenberg collection
\cite{Gutenberg} of texts, with $L=10000$. This word it is
practically impossible to be used as a keyword and therefore we
can assume that the distribution would be simply Binomial.
However, it is clear that the distribution is not Binomial; it is
highly skewed and far away from the Binomial distribution with
that frequency \cite{truam}.

 Empirically we have found that the distribution is Gamma distribution for all the words if the different meanings of
 the homonyms are regarded as different words.

 By definition, the Gamma distribution is:
\begin{equation}\label{gamamdist}
P(x;w)= e^{-x b}x^{a-1}b^a /\Gamma(a),
 \end{equation}
where $b$ is a parameter independent of the length of the text
e.g. it depends only on the word and the class of text we are
regarding.  The parameter $a$ is proportional to the length of the
text $L$.

 The empirical proof of the statement about the Gamma distribution can be performed on a text corpus with sufficiently
 large size, dividing it in small fragments.  These segments must be chosen with a sufficient length $L$ in order to
 have $L p_w\gg 1$, where $p_w$ is the probability of occurrence of the word $w$.

 We have checked the above hypothesis of Gamma distribution on the British National Corpus (BNC) \cite{BNC} and on a set of about
 19000 English texts chosen from the Gutenberg collection and we found an excellent agreement ($p>0.8$) with the
 experimental data for all the words with $p_w>5/10000$ \cite{truam}.

The statement that the distribution of a given word is Gamma is not common in the literature. In this article we do
not give a model to explain it. However independently of the nature of the underlying process we found that the Gamma
distribution fits well the empirical data.

% Some simple intuition about the Gamma distribution can be given, supposing that the language confirm some
%probabilistic grammar. Then fixing some word because of the recursive nature of the language the probability of
%the word will be exponentially falling. The sum of exponentially distributed variables is by definition Gamma distributed.
%However, the argument is not strict.

\begin{figure}[t]
\begin{center}
\begin{minipage}{7cm}
\epsfxsize 7cm %\epsfysize=5cm
\epsfbox{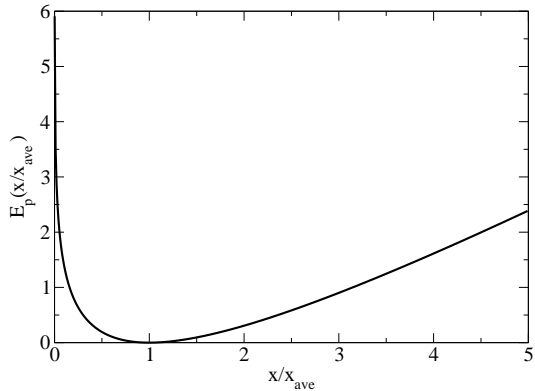}
\end{minipage}
\caption{\small Potential energy of a word
according to the number of occurrences. It consists of two parts
-- the logarithmic falling part varying for values of the argument
from zero to the mean frequency of the word and a linear
increasing part, predominant at the range where the frequency of
the word is larger than its mean frequency in the language.
}\label{potenergy}
\end{center}
 \end{figure}

 Further, we have analyzed the asymptotic behavior of the distribution. To achieve this, we replicated the text $s$ times
 and consideed the limit  $\lim_{s\rightarrow\infty}[\log P(s x; w; s a,b)]/s=a-b x-a \log a + a \log x + a \log b$.
Using that the mean of $x$ is $\bar x=a b$, we obtained for the
asymptotic behavior of $\log P(x)$ the following final expression:
\begin{equation}\label{Ep}
E_p(x;w)=-\log P(x)=-\bar x b \left[ 1-\frac{x}{\bar x}+
\log\left(\frac{x}{\bar x}\right)\right].
 \end{equation}
$E_p$ can be regarded as a potential energy of the word $w$ in the
language. The logarithmic member corresponds to the entopic part
of the energy\cite{chaos}, while the linear one accounts for the
excess of words of a given type in the text.  A normalized energy
curve is given in Fig.~\ref{potenergy}.

%% file: 04_free_energy.tex
\section{The free energy}

 Using the above considerations, the corresponding partition function for a given word $w$ is:
\begin{equation}\label{Partfunc}
Z(w, \beta)=\sum_{m=1}^{N_t} G(m, N_t)) \exp(-\beta E_p(m, N_t)),
 \end{equation}
where we have used the argument that the energy for a single word is given by its potential energy Eq.(\ref{Ep}).

 Introducing in the above equation the expression for the number of degenerations  $G(m, N_t) = {{N_t}\choose {m}}$ and identifying
 the parameters $\bar x = N_v$ and $x=m$, we arrive to the following expression for the partition function:
\begin{equation}\label{z1}
Z(w, \beta) = \sum_{m=1}^{N_t} \exp(- \beta E_{tot}(m, N_t)).
 \end{equation}
Here
\begin{eqnarray}\label{z2}
   E_{tot}(m, N_t)= - \frac{1}{\beta} \log {{N_t}\choose {m}} + \nonumber\\
      N_v b \left[ 1-\frac{m}{N_v}+
   \log\left(\frac{m}{N_v}\right)\right]
\end{eqnarray}
is the total energy corresponding to some word $w$ and we have introduced the degeneration factor inside the exponent.

 As can be seen, the total energy for one word is composed by a potential part $E_p$ and by a combinatorial part  $\frac{1}{\beta}\log G(m, N_t)$.

 Finally, the full free energy of the text is a sum over all the words of the text:
\begin{equation}
F(\beta) = - \frac{1}{\beta} \sum_{w} \log Z(w,\beta).
 \end{equation}

 The equation for the order parameter $m$ can be obtained by using the saddle-point method and the Stirling approximation,
 $\log N! \approx N \log N - N$, $N\gg 1$:
\begin{equation}\label{m1}
\frac{d F}{dm} =
\frac{1}{\beta}\log{\frac{m}{N_t-m}}+b\frac{N_v-m}{m}=0.
 \end{equation} This equation can be solved in a closed form giving the following final expression:
\begin{equation}\label{m2}
m = N_t\frac{b \beta N_v/N_t}
           {b \beta N_v /N_t
             + W(b \beta N_v/N_t\ e^{b \beta - b \beta N_v/N_t})},
 \end{equation}
where $W(.)$ is the Lambert W function \cite{specfunc}. The ratio $m/M_t$ is a monotonously increasing function of $\beta$ and $N_v/N_t$.

 For small values of the $N_v/N_t$, the ratio $m/N_t$ is small for any temperature, growing later above some critical value of $N_v/N_t$.

 Further we can consider the rest of the thermodynamic quantities. The entropy $S$ for a single word is:
\begin{eqnarray}\label{entrop}
S \equiv - \frac{\partial F}{\partial T} = N_t \log N_t -m \log m -
\nonumber\\
(N_t-m) \log (N_t - m).
 \end{eqnarray}
Note that Eq.(\ref{entrop}) approaches asymptotically the "usual"
entropy $-m\log m$ for $N_t\rightarrow \infty$. However, when $m$
is of order $N_t$, this is not longer true. Substituting
Eq.(\ref{m2}) in Eq.(\ref{entrop}), we obtain explicit expression
for the entropy as a function of the systems parameters. It is
monotonously decreasing function of $\beta$ and $N_v/N_t$.

\begin{figure}[t]

\begin{center}
\begin{minipage}{7cm}
\epsfxsize 7cm %\epsfysize=5cm
\epsfbox{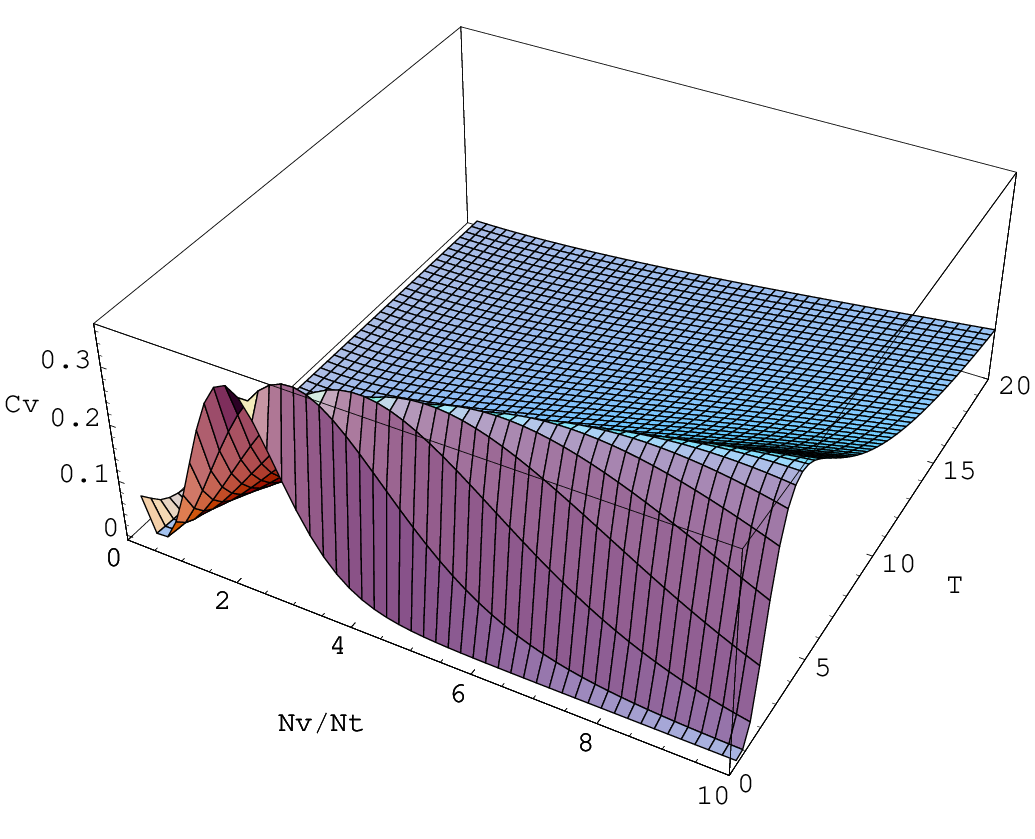}
\end{minipage}
\hfill
\begin{minipage}{6cm}
\epsfxsize 6cm %\epsfysize=5cm
\epsfbox{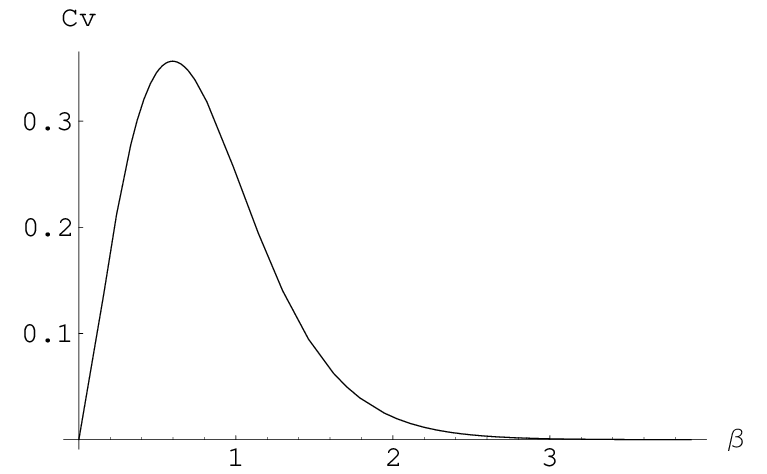}
\end{minipage}
\caption{\small (Color online) The ``specific heat'' $C_V$.
}\label{figcv}
\end{center}
 \end{figure}

 The second derivative of the free energy is related to the ``specific heat''
 (Fig.~\ref{figcv}): \[ C_V = -T \left(\frac{\partial ^2 F}{\partial T^2}\right)_V. \]
 In the context of the statistical model of texts, this quantity can be interpreted in
 the following way: if $C_V$ for a given word is high, then replacing this word by another one,
 or omitting it, will introduce relatively big distortion in the text, leading to significant change
 of the total energy. On the other had, replacing word with negligible $C_V$, will have no relevant consequence on the text.

%  The behavior of the specific heat shows a maximum for some value of the temperature.  We have tested this behavior on several lengths of texts in order to see any size effects and   we have found that the behavior is independent on the size.

 We use the usual notation for $C_V$, adopted in thermodynamics for isochoric process, where the
 volume of the system is fixed,  although what is fixed in this consideration is the number of
 occurrence  for a given word. We also represent a section of Fig.~\ref{figcv} for $b=1, N_v=5$ in the lower panel of the figure.

\begin{figure}[t]
\begin{center}
\begin{minipage}{6cm}
\epsfxsize 6cm %\epsfysize=5cm
\epsfbox{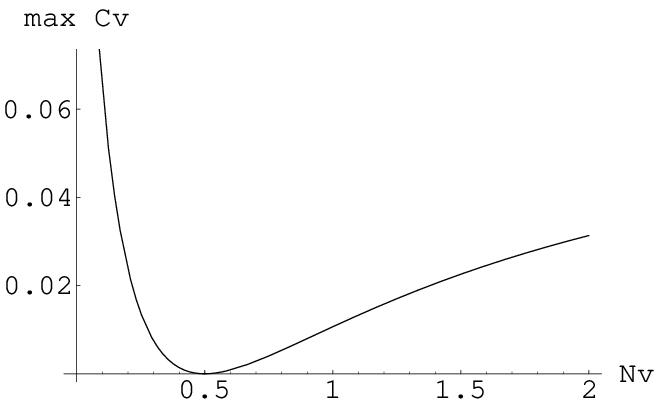}
\end{minipage}
\hfill
\begin{minipage}{6cm}
\epsfxsize 6cm %\epsfysize=5cm
\epsfbox{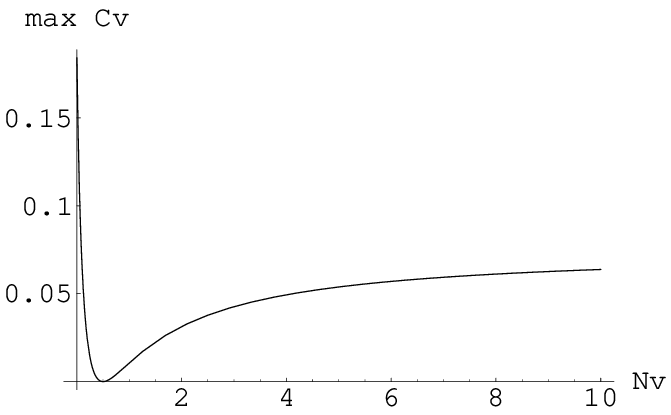}
\end{minipage}
\caption{\small The maximal values of $C_V$ as a
function of $b N_v/N_t$. The upper panel is a zoomed version of
the left one. }\label{cv1}
\end{center}
 \end{figure}

As can be seen, $C_V$ starts form zero at $T=0$, then expresses  a
maximum for  some temperature $T$, after which it further
decreases to zero. The temperature corresponding to the maximum of
$C_V$ is easy to be exploited numerically. It is $T_{max}=2.4 b
N_v/N_t+ 1.043$ and it is linear
 with respect to $N_v/N_t$. The maximum value of $C_V$ as a function of the parameter $ b N_v/N_t$ is represented in Fig.~\ref{cv1}.

We have tested numerically the dependence of the position and the
hight of the maximum of the specific heat on several lengths of
texts in order to see any size effects and we have found that the
behavior is independent on the size \cite{fss}.

Using similar approach for images in the thermodynamic limit
\cite{Bialek}, i.e. when the size of the blocks goes to infinity,
one expects a divergence of the specific heat \cite{criticality}.
This is due to the fact that in images one can find a homogeneous
statistics for different resolutions and image sizes and both can
go to infinity. However, similar behavior is not observable in the
case of texts, because a single text that explains a given concept
has a rather limited size and a finite "resolution", and cannot be
extended. That is why in our model for statistical mechanics of
written texts, considering the words as independent, one only
observes smeared behavior of the specific heat parameter.

% In the next Section we will discuss in details that the  separation of the words of a given text in three  classes: function words, keywords and common words, gives different  specific heat for each class of the words\footnote{note that a-priory we do not introduce classes as an attribute of the word}.  Namely, the function words, although they are much more frequent, have relatively low specific heat $C_V$, which peak is located within the low-temperature region. This means that in general the function words are almost always in  equilibrium with the thermostat and a lot of "noise" is needed to change that configuration,  although each component is rather volatile. The common words have higher temperature of maxima of the $C_v$, but lower specific heat. That means that these words can be easily  substituted by other words -- mostly synonyms, in order to avoid repetitions. The keywords show very high values of their specific heat behavior.  They are either difficult to be substitute with other words or are much more frequent in the text than in the language in general, that characterizes  them as very "high energy" units.

%% file: 05_numerical.tex
\section{Numerical experiments}

 To check the above results experimentally on real texts, we used several corpora of texts. First, we used a BNC corpus, as a
 standard and equilibrated corpus of English texts with some $10^8$ words.  Second, we used a collection of about 19000 English texts of
 the Gutenberg collection (GC) with size $5.10^7$ words.  To check specific domains we used single articles, as well as a collection of 500 articles
 from the non-linear physics archive (NL) offered by the xxx.archiv.gov repository. In order to avoid problems with the different versions of the
 articles, we used only the first version of each article. Also, we used a list of 257 closed-class words of English instead of the function words.

 For estimating the parameters $a$ and $b$ of the Gamma distribution of a single word, we used BNC and GC that give practically the same results.
 The parameter $b$ is within the range 0.01-20 with an average value 0.25 and the parameter $a$ belongs to the interval from 0 to 2.6 for a length of the
 words $L=10000$. Note that the parameters $a$ and $b$ are well defined and with a sufficient confidence only if $p_w L\gg 1$, where for all
 practical purposes we can suppose that $5\gg 1$. Thus, within the corpus of $10^8$ words, the parameters are well defined for less then 2400 words.
 For the rest of the words we used some simplifying assumption due to the difficulty to prove or disprove reliably a hypothesis with two degrees of
 freedom ($a$ and $b$) having less then five measures for their estimation.

 The hypothesis we have adopted was that the less frequent words have the same value of the parameter b for all the words.  In this way we could join
 all the words that are not frequent enough for estimating that parameter.  The results are very close to the mean vale of $b$.  The parameter $a$,
 being proportional to the length of the text, is not so critical to estimate  (actually we need only $N_v$ and $b$).

\begin{figure}[t]
\begin{center}
\begin{minipage}{6.70cm}
\epsfxsize 6.7cm %\epsfysize=5cm
\epsfbox{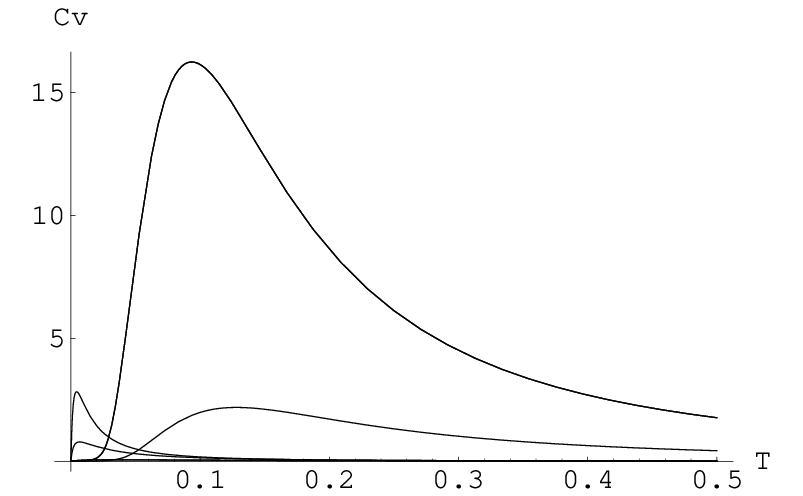}
\end{minipage}
\hfill
\begin{minipage}{6.70cm}
\epsfxsize 6.7cm %\epsfysize=5cm
\epsfbox{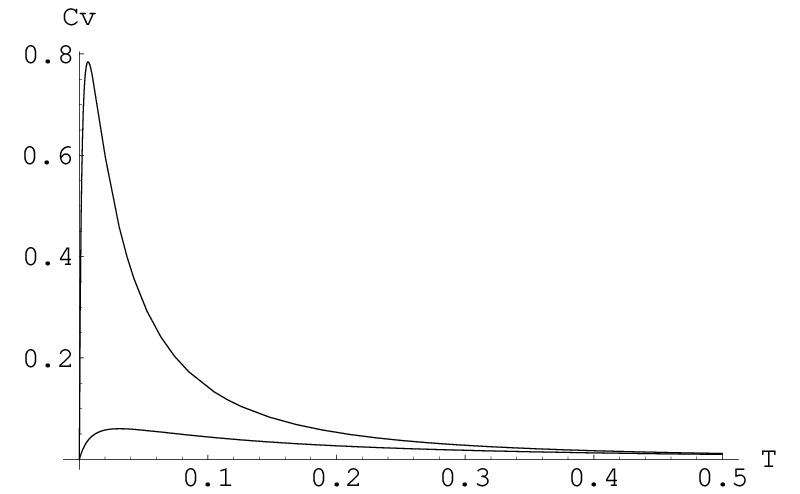}
\end{minipage}
\caption{\small $C_V$ for different words of one
and the same text. The upper two curves of the left panel
represent two keywords of a given text (``topology'' and
``topological''). The lower curves of the left panel represent two
functional words (``the'' and ``are''). On the lower panel the
curve of ``are'' is zoomed in order to represent also the typical
common word ``important''. }\label{cvt}
\end{center}
 \end{figure}

 We expected domain nonspecific behavior of the function and the common words, and domain and text specific behavior of the keywords.

 Figs.~\ref{cvt} show a typical behavior of $C_V$ for keywords (the two upper curves in the upper panel), for function words
 (the two curves upper-down in the same panel) and for common words (the lower curve in the lower panel).

 As the function words have much higher frequency of occurrence, one can expect that they will have predominant role in the specific heat.
 However this is not observed. The specific heat for the keywords is much higher than the corresponding one for the function words.
 Even smaller specific heat is carried by the common words.

 These results can be interpreted as an indication than the most vulnerable speech parts are the common words and the most resistant ones are
 the domain-specific (keywords).
 %Also it can be noted that the temperature range of the corresponding maxima is different.

 Alternatively, one can interpret the temperature factor as a weight of the combinatorial term that depends only on the text. Thus, it is not
 surprising that the language dependent part (the function words) shows the maximal $C_V$ at lower temperature (see Eq. \ref{z2}).
 On the contrary, the keywords in the text, which are not so language dependent, have the maximum of $C_V$ for higher temperatures.

\begin{figure}[t]
\begin{center}
\begin{minipage}{7.40cm}
\epsfxsize 7.4cm %\epsfysize=5cm
\epsfbox{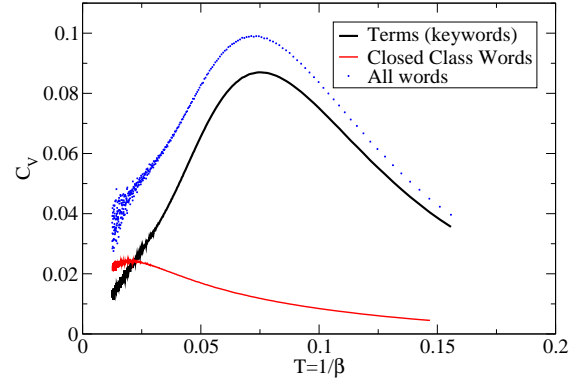}
\end{minipage}
\caption{\small (Color online) Experimentally measured $C_V$ on a
single text. The part of the $C_V$ corresponding to the common
words is very small to be shown in that scale. }\label{Cvtotal}
\end{center}
 \end{figure}

 Considering all the words and having the parameters $\bar x$ and $b$ for each of them, we can calculate numerically the free energy $F$, the
 entropy $S$ and the specific heat $C_V$ for the whole text. The result for $C_V$ is shown in Fig.~\ref{Cvtotal}.

 What is observed experimentally is the lack of a well pronounces maxima of $C_V$ for the function words, less expressed maxima for the common
 words and well pronounces maxima for the keywords.  The function words express the structure of the language, e.g. represent its grammar.
 The keywords, on the other hand, are expressions of the semantic and the pragmatic structure of the text.  If we represent that structure as a
 semantic graph, similar to \cite{semgraph3}, we can suppose that the keywords reflect the structure of that graph independently of the grammar
 or the language we chose to express it as a text.

 According to Fig. \ref{Cvtotal} we can observe that there is a wide temperature range between the maximum of $C_V$ corresponding to function
 words and the maxima of the keywords. Within this area we can expect that the solution will contain few function words but the rest of the words
 will be sufficient for the interpretation of the text.

 In order to check that, we took an abstract of a given article and deleted the deposited words with a probability $m/N_t$.  The result is shown in the
 following boxes, where we represented the same text for different values of the temperature.  The over-stroked words are chosen by their probability
 of deposition.  It can be seen that the method extracts very well the meaning and ignores the language structure.  The extraction is perfect in the
 last box, where the temperature is lower. Note that the words are represented only by their parameters $\bar x$ and $b$.  The program has no
 notion of ``function word'' ``common word'' or ``keywords''.

 \input{abstracts.tex}

%% file: abstracts.tex
 {\small --------------------------------------------------\\ \xout{\sout{Abstract}} . \xout{\sout{Following}}
 Gardner , we \xout{\sout{calculate}} the \xout{\sout{information}} \xout{\sout{capacity}} \xout{\sout{and}} \xout{\sout{other}}
 \xout{\sout{phase}} transition \xout{\sout{related}} parameters \xout{\sout{for}} a symmetric \xout{\sout{Hebb}} \xout{\sout{network}}
 with small world \xout{\sout{topology}} \xout{\sout{in}} mean - field \xout{\sout{approximation}} . \xout{\sout{It}} was found that the
 topology dependence \xout{\sout{can}} be described \xout{\sout{by}} \xout{\sout{very}} small \xout{\sout{number}} of parameters ,
 namely \xout{\sout{the}} \xout{\sout{probability}} of existence \xout{\sout{of}} loops \xout{\sout{with}} \xout{\sout{given}}
 length . \xout{\sout{In}} \xout{\sout{the}} \xout{\sout{case}} \xout{\sout{of}} \xout{\sout{small}} \xout{\sout{world}} topology ,
 closed \xout{\sout{algebraic}} \xout{\sout{set}} \xout{\sout{of}} \xout{\sout{equations}} with \xout{\sout{only}} \xout{\sout{three}}
 parameters was \xout{\sout{found}} \xout{\sout{that}} \xout{\sout{is}} easily \xout{\sout{to}} be solved. \vskip1mm {1. T=100.
 In this case the temperature is very high, which makes the text almost unreadable.} \vskip0.3cm --------------------------------------------------\\
 \xout{\sout{Abstract}}. \xout{\sout{Following}} Gardner, we calculate the \xout{\sout{information}} capacity \xout{\sout{and}} \xout{\sout{other}}
 phase transition \xout{\sout{related}} parameters \xout{\sout{for}} a symmetric Hebb \xout{\sout{network}} \xout{\sout{with}}
 small world \xout{\sout{topology}} \xout{\sout{in}} mean - field \xout{\sout{approximation}} . \xout{\sout{It}} was found that the
 topology dependence \xout{\sout{can}} be described \xout{\sout{by}} \xout{\sout{very}} small \xout{\sout{number}} of parameters ,
 namely \xout{\sout{the}} \xout{\sout{probability}} of existence \xout{\sout{of}} loops \xout{\sout{with}} given length .
 \xout{\sout{In}} \xout{\sout{the}} \xout{\sout{case}} \xout{\sout{of}} \xout{\sout{small}} world topology , closed
 algebraic \xout{\sout{set}} \xout{\sout{of}} equations with \xout{\sout{only}} \xout{\sout{three}} parameters was
 \xout{\sout{found}} \xout{\sout{that}} \xout{\sout{is}} easily \xout{\sout{to}} be solved . \vskip1mm {2. T=0.167.
 This version of the text corresponds to values of the parameters that correspond to the region located after the peak of the "specific heat" $C_V$,
 corresponding to the keywords.} \vskip3mm --------------------------------------------------\\ \xout{\sout{Abstract}}.
 Following Gardner, \xout{\sout{we}} calculate the \xout{\sout{information}} capacity \xout{\sout{and}} \xout{\sout{other}}
 phase transition \xout{\sout{related}} parameters \xout{\sout{for}} a symmetric Hebb \xout{\sout{network}} \xout{\sout{with}}
 small world topology \xout{\sout{in}} mean - field approximation . \xout{\sout{It}} \xout{\sout{was}} found that the topology
 dependence \xout{\sout{can}} be described \xout{\sout{by}} \xout{\sout{very}} small \xout{\sout{number}} of parameters ,
 namely \xout{\sout{the}} probability of existence \xout{\sout{of}} loops \xout{\sout{with}} given length .
 \xout{\sout{In}} \xout{\sout{the}} \xout{\sout{case}} \xout{\sout{of}} small world topology , closed algebraic
 \xout{\sout{set}} \xout{\sout{of}} equations with \xout{\sout{only}} \xout{\sout{three}} parameters was \xout{\sout{found}}
 \xout{\sout{that}} \xout{\sout{is}} easily \xout{\sout{to}} be solved. \vskip1mm {3. T=0.05. Extraction of the text for values
 of the parameters that correspond to the region located between the peaks of the "specific heat" $C_V$ corresponding to the function words and the
 keywords}.
 \vskip0.5cm \vskip2pt --------------------------------------------------\\ \xout{\sout{Abstract}}. Following Gardner, \xout{\sout{we}}
 calculate the information capacity \xout{\sout{and}} \xout{\sout{other}} phase transition related parameters \xout{\sout{for}}
 a symmetric Hebb network \xout{\sout{with}} small world topology \xout{\sout{in}} mean - field approximation . \xout{\sout{It}} \xout{\sout{was}}
 found that the topology dependence \xout{\sout{can}} be described \xout{\sout{by}} \xout{\sout{very}} small number of parameters ,
 namely \xout{\sout{the}} probability of existence \xout{\sout{of}} loops \xout{\sout{with}} given length .
 \xout{\sout{In}} \xout{\sout{the}} \xout{\sout{case}} \xout{\sout{of}} small world topology , closed algebraic
 \xout{\sout{set}} \xout{\sout{of}} equations with \xout{\sout{only}} \xout{\sout{three}}
 parameters \xout{\sout{was}} \xout{\sout{found}} \xout{\sout{that}} \xout{\sout{is}}
 easily \xout{\sout{to}} \xout{\sout{be}} solved. \vskip1mm {4. T=0.0125. The extraction of the text corresponding to values of the
 parameters that correspond to the region located it in the vicinity of the peak of the "specific heat" $C_V$, corresponding to the function words.}
 \vskip2pt --------------------------------------------------\\ }

%% file: 06_application.tex
\section{Toy Application}

\begin{figure}[t!h]
\begin{center}
\begin{minipage}{7cm}
\epsfxsize 7cm %\epsfysize=5cm
\epsfbox{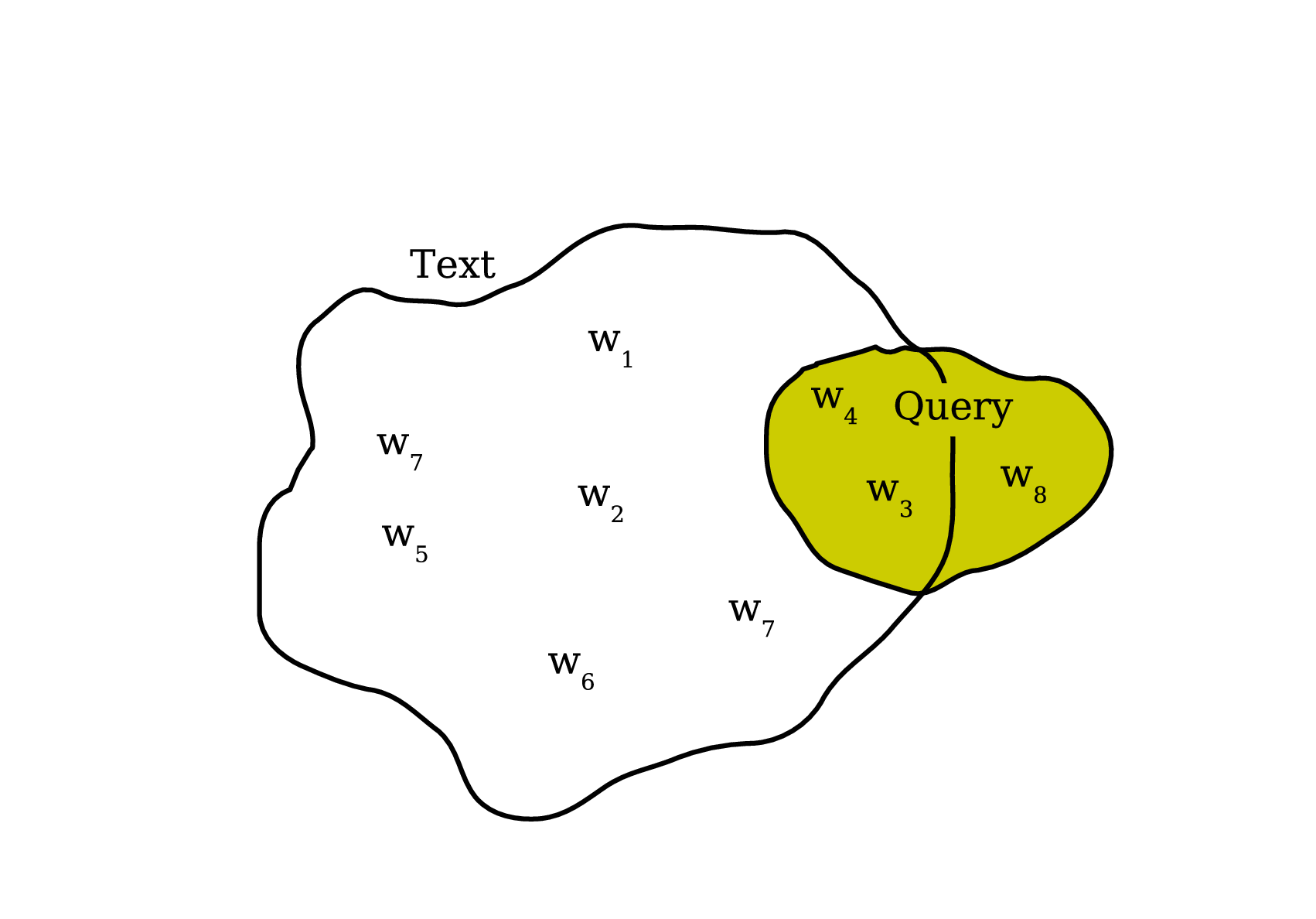}
\end{minipage}
\caption{\small (Color online) We regard the query as a subset of
words. The effective energy will give as a measure of the
relevance of the query. $Prob(Q)\propto exp(-E(w_3)/kT-E(w_4)/kT)$
}\label{fig_app}
\end{center}
 \end{figure}

Let us consider the above consideration in the following context: we
consider a text and we are asking to what extend certain word is characteristics for the text.

If the word is a keyword, it is of course characteristics for the
text. A typical information retrieval application tries to use
these kinds of words in order to extract a text relevant to some
query. The probability to have $m$ times some word $w$  relative
to the mean frequency of this word in the language is  $\propto
exp(-E_{tot}^{(w)}(m,\beta)$, (see Eq.\ref{z2}). Then, for a set
of different words as independent entities,  the probability to
have this set of words in the text would be $\propto exp(-\sum_w
E_{tot}^{(w)}(m,\beta))$.

 If we ask whether some set of words $Q\equiv\{w_{q1},w_{q2},...,w_{qm}\}$ are relevant to the text, and if relevant is considered as much more
 probable that its average use in the language, then $Q$ is relevant if the energy of the words forming $Q$ is high.  If some words occur in $Q$ and
 not in the text, then the Gibbs multiplier will be zero and this word will be ignored.

 The concept is very easy to implement. Just calculate the effective energy of the words in a text and store them as pairs $(w,E^{(w)}(T))$ for several
 temperatures. Then using the query, one can sum the energies of the words  (see Fig. \ref{fig_app}). According to the present theory, the quality of the
 result of the query does not depend on the length of the text and the query.

 The query can perform better than our model which assumes the independence of the words in the text and considers the query and the text as a
 set of words.  This model is close to the vector information retrieval model \cite{nlp_ir}, but it is richer, because the Gamma distribution is
 bi-parametric one. 

Query performance on a real implementation is currently under evaluation using strict IR
criteria and future results will be published elsewhere.

%% file: 07_discussion.tex
\section{Discussion and future directions}
As has been shown from the above results, the statistical mechanics approach permits a relatively easy theoretical analysis and a very fast
simulation procedure, which make it promising.

The method has some advantages in comparison with the usual IR
methods. First of all the queries correspond to the real
probability measures conditioned to the language. There is no
empirical moment of choice. Second, it is relatively easy to
introduce a interaction between the words, e.g. to introduce
conditional probabilities that goes beyond simple bi-grammar
models \cite{bigram}. Because the stable bi-grams are much more
frequent in one and the same text than throughout the corpus, it
is logical to suppose that the interactions are week. If one
introduces them as a perturbation of the energy, the resulting
model can be very resistant to errors and on the same time can
respect the language structure.

As a further step, we can consider different modifications of the model proposed in this article. For example, the potential energy, derived
experimentally and corresponding to the frequencies of the words in texts with fixed length can be substituted by different functions seeking
different characteristics of the text.

In this paper we use the words as a convenient starting point. However the approach is not limited to words. Another interesting choice is the use
of maximal common prefixes, e.g. the strings of the texts with maximal length that coincide.
%The techniques seem to be computationally demanding,
%but by ordering the rotations of the vocabulary and the text and by merging them, it is possible to do it with $L \log L$ operations and with a
%memory of the order of  $c L$. It is possible to do it also incrementally.

 Allowing only non-overlapping strings and conditioning the text to itself, the number of ``molecules'' in $T = 1$ would be the length of LZ compressed
 file and therefore resembles its Kolmogorov complexity \cite{Kolmogorov}.  Distances similar to that used by \cite{zurek, sundresh} can be easily
 calculated introducing a chemical potential. The disadvantage of these distances is that they are not operational with short texts and keywords.
 Thus although they give best results in tasks measuring proximity of texts, they are difficult to use for information retrieval purposes. This is
 due to the extremely sparse representation needed in order to compress the text.

 The fact that this type of distances can be regarded as an extreme case, gives us the ground to expect that the behavior of the system would be richer
 within the finite temperature range.

 Using overlapping strings and conditioning not only between two texts, but also between the language, the knowledge area, the author and similar
 characteristics, can give much denser representation and could lead to very interesting information retrieval applications.

%% file: 08_relatedwork.tex
\section{Related work}

The problem of keyword detection starts with the seminal work by
H. P. Luhn \cite{Luhn} in which he uses statistical information derived from
word frequency and distribution to compute a relative measure
of significance, first for individual words and then for sentences with
the purpose of outputting an ``auto-abstract'' of the article in question. 

In last recent years, this field has been addressed with 
Statistical Mechanics tools. The contributions \cite{Herrera, Ortuno, Zhou} focus
on statistical information referring to the spatial use of the words
in human written texts as opposed to the spatial distribution of words
in random shuffled texts. They argue that the spatial deviation of
the distances between successive occurrences of a word is an excellent
parameter to quantify its relevance for the text, i.e. keywords tend to form clusters, while function words are essentially uniformly spread in a text.

A work of different nature, but also in the statistics field, has recently shown that, by considering words as a network
of interacting letters, the maximum entropy models, which are
consistent with pairwise correlations among letters, provide a
very good approximation to the full statistics of four letter
words \cite{4letters}.

Using a similar approach to the one presented in this article, but in
the context of image analysis, a statistical mechanics formulation has been defined
 for the distribution of small image patches
\cite{Bialek}. By assuming Boltzmann distribution of the patches,
the authors derived the entropy and the heat capacity and showed
that the behavior of the heat capacity is divergent in the vicinity
of the critical temperature.

%% file: 09_conclusion.tex
 \section{Conclusion}

 In the present article we propose a statistical physics approach for the analysis of human written text.
 By introducing the concept of energy of interaction between the text and the corpus (the language), and taking into consideration the realistic
 distribution of the words
 inside a given large text corpus, we are able to derive the thermodynamics parameters of the system in a closed analytical form.

 The behavior of the specific heat of the system is different for different kinds of words  (keywords, function words and common
 words).
 It is universal and independent on the selected text.  We also show that the temperature range, where the maxima for these types
 of words occur, is different.

 Finally we discuss a possible application of the method in order to construct queries on text database.
 Thus, without having knowledge of the text, we could judge about the structure and the functionality of the different parts of the text,
 which could be useful for several information retrieval applications.

%% file: molec.bbl
\begin{thebibliography}{99}
\bibitem{footnote1} Here we will not consider texts like the
genome's sequence, computer logs, multi-language texts and similar.
\bibitem{footnote2} Halomorphemes are the words that  occur only once in   a text.
\bibitem{bigram} Bi-grammar modls are models in which one analyses
the distribution of two consecutive words (bigrams).
\end{thebibliography}
